\documentclass{aa}

\usepackage{natbib}
\usepackage[english]{babel}
\usepackage{amsmath}
\usepackage{amssymb}
\usepackage{graphicx}
\usepackage{xcolor}
\usepackage[colorlinks,citecolor=blue,urlcolor=blue,bookmarks=false,hypertexnames=true]{hyperref}
\usepackage[export]{adjustbox}

\bibpunct{(}{)}{;}{a}{}{,} 

\LARGE  \title{Evolution of quasi-periodic eruptions in the post-tidal disruption event accretion disk perturbed by an orbiting star}

   \author{Martin Mondek \inst{1}
          \and
          Michal Zajaček \inst{1}
          \and
          Henry Best \inst{1}
          \and
          Taj Jankovič \inst{3}
          \and
          Vladim\'ir Karas \inst{2}
          \and
          Petr Kurfürst \inst{1}
          }

   \institute{Department of Theoretical Physics and Astrophysics, Faculty of Science, Masaryk University, Kotlářská 2, 611 37 Brno, Czechia
   \and
     Astronomical Institute of the Czech Academy of Sciences, Boční II 1401, 141 00 Prague, Czechia
    \and
    Institute of Physics of the Czech Academy of Sciences, Na Slovance 1999/2,182 21 Praha 8, Prague, Czechia
    }
  \date{accepted March 27, 2026}

\begin{document}

\abstract
{Quasi-periodic eruptions (QPEs) are a recently discovered class of highly variable X-ray bursts originating in galactic nuclei. These high-amplitude bursts exhibit a periodicity ranging from tens of minutes to several days and they are also characterized by variable peak amplitudes that can vary by a factor of few. While multiple physical models have been proposed to explain QPE light curves, none of them can fully account for all the observed features. A possible connection between QPEs and tidal disruption events (TDEs) has been suggested, particularly due to past optical/UV outbursts that can be traced back for several sources, the long-term decay in the continuum luminosity, and the soft, thermal-dominated X-ray spectrum.}
{Our primary goal is to verify whether the long-term decrease in eruption amplitudes detected for some QPE sources is consistent with a scenario where the accretion disk had been formed following a TDE.}
{In this work, we adopted a simplified extreme mass ratio inspiral (EMRI) scenario, where a  solar-type star orbits a supermassive black hole (SMBH) and collides with an accretion disk twice per orbit, generating eruptions. We assumed a post-TDE disk that follows a temporal power-law decline in mass accretion ($\propto t^{-p}$, $p>0$). As our aim is to develop a toy-model scenario, we  used purely analytical methods, without considering all intervening processes in their full scope.}
{Our results indicate  that (i) the observed long-term decline in QPE amplitudes can be reproduced if the first monitored epoch occurs years to a few decades after the tidal disruption; and (ii) stellar mass loss caused by ablation can play an important role in the evolution of QPE amplitudes in systems with heavy main sequence stars.}
{}

\titlerunning{Evolution of QPEs}
\maketitle

\nolinenumbers

\section{Introduction}
\label{sec1}
\indent

The last decade of observations of galactic nuclei (GNs) has revealed environments that can be studied both observationally over wide spectral, temporal, and spatial ranges and within the theoretical framework of general relativity as well as beyond it \citep[see e.g.,][for reviews]{2017FoPh...47..553E,2022RvMP...94b0501G,2024A&ARv..32....3G}. GNs are expected to contain supermassive black holes (hereafter, SMBHs) and based on their activity (i.e., their variable energy output), we can roughly divide them into active galactic nuclei (AGNs) with much higher energy outputs than the second category, constituting the majority in the Local Universe: quiescent GNs.

The dynamics of GNs is difficult to probe observationally due to the relatively small sizes of their central engines, making them unresolvable except for a few nearby cases \citep{EHT2,EHT1,2024A&A...684A.167G}. In some rare cases, however, there are recurring or almost periodic mechanisms that help us unravel their nature. We refer to those as repeating nuclear transients \citep[RNTs; see][for a review]{2024arXiv241104592S} and GNs exhibiting them have a potential to serve as natural laboratories of accretion and gravitational physics. These transient nuclear phenomena occur repeatedly; hence, various dynamical models can be verified over longer timescales of weeks to months, up to even years.

X-ray quasi-periodic eruptions (QPEs) are relatively new, recurrent nuclear phenomena that form a subclass of RNTs towards the lower periodicity end \citep[for a review, see][]{kara2025}. They are characterized by fast bursts in the soft X-ray band that repeat roughly every few hours to a few days \citep{2023NatAs...7.1368E}. Hence, if they are associated with the orbital period corresponding to the semimajor axis, $a$ (two QPEs per orbital period), they have the potential to probe the inner accretion flow around the SMBH on a length scale of
\begin{equation}
    \frac{a}{R_{\rm g}}\sim 60 \left(\frac{P_{\rm QPE}}{2\,\text{hours}} \right)^{2/3} \left(\frac{M_{\bullet}}{10^6\,{\rm M}_{\odot}} \right)^{-2/3}\,,
    \label{eq_radius_QPE_periodicity_mass}
\end{equation}
where $R_{\rm g}=GM_{\bullet}/c^2\sim 1.5 \times 10^{11}(M_{\bullet}/10^6\,{\rm M}_{\odot})\,{\rm cm}\sim 2.12\,(M_{\bullet}/10^6\,{\rm M}_{\odot})R_{\odot}$ as the gravitational radius of the SMBH.
During these flares, the X-ray count rate increases by one to two orders of magnitude relative to the quiescent level that presumably originates in the accretion disk alone. The burst phases of QPEs have thermal-like spectra with $kT\approx 100$--$250\,\text{eV}$ ($k$ is the Boltzmann constant, $T$ is temperature), while the quiescent emission is near $kT\approx 50$--$80\,\text{eV}$ \citep{Frenchie}. The duty cycle, defined as the ratio of the mean eruption duration to the mean recurrence time, is typically $\sim 10-30\,\%$.
The peak X-ray luminosity remains in the range $\approx 10^{41}-10^{43}\,\text{erg}\,\text{s}^{-1}$.
GSN 069 \citep{2019Natur.573..381M} was the first source detected, followed by other QPEs in the nuclei of the following host galaxies: RX J1301.9$+$2747 \citep{2020A&A...636L...2G}, eRO-QPE1, and eRO-QPE2 \citep{2021Natur.592..704A}, XMMSL1 J024916.6-04124 \citep[only 1.5 bursts have been detected;][]{2021ApJ}, eRO-QPE3, eRO-QPE4 \citep{arcodia2024}, and the most recent: eRO-QPE5 \citep{arcodia2025} and J2344 \citep{2026A&A...706L..15B}.

Observational analyses have also indicated that some QPEs are associated with recent TDE hosts. The first QPE source, GSN 069 exhibits signatures of a post-TDE evolution in the long-term light curves \citep{2023A&A...670A..93M}. The QPEs were also detected in the ``Ansky'' (i.e., nickname for the SMBH in question; \citealt{Ansky}, \citealt{2025ApJZhu}), which was the site of an untypical TDE. Periodic flaring was also confirmed in AT2019qiz \citep{nicholl2024}. After the optical TDE peak, both of these TDE sources exhibited QPEs in the soft X-ray domain, while their optical flux declined as a power law. The most recent source, AT2022upj \citep{chakraborty2025}, started exhibiting eruptions 2 years after a TDE and has been associated with the emission of extreme coronal lines, together with AT2019qiz. Finally, a hint toward QPEs was observed in ``Tormund" \citep{Tormund}. For a statistical overview of the key characteristics, we present four graphs of the most relevant QPE properties: the QPE recurrence times, durations, peak luminosities, and SMBH masses in Appendix~\ref{apendixa}.

The QPE class has not yet been strictly defined because the population sample is not large enough. However, the approximate set of criteria for a RNT to join this group  \citep{Miniutti_2023}   follows:
\begin{enumerate}
    \label{conditions}
    \item a relative increase in the soft X-ray count rate compared to the quiescent state with a ratio of 10--100;
    \item  peak soft X-ray luminosities of the eruptions  in the range $ 10^{41}\text{--}10^{43}\,\text{erg}\,\text{s}^{-1}$;
    \item low-mass host galaxies with $\log{(M_\text{galaxy}/M_{\odot})} \approx 9\text{--}10$,  implying lower-mass central black holes;
    \item no significant UV, optical, or IR periodicity, namely, their variability is not correlated with the eruptions;
    \item eruption recurrence times in the range of hours to days;
    \item a closed hysteresis cycle in the $L-kT$ plane, namely, the temperature evolution colder $\rightarrow$ warmer $\rightarrow$ colder during eruptions.
\end{enumerate}

An example of an RNT that meets several of these criteria is Swift J0230$+$28; however, it cannot be fully classified as a QPE source. Although its peak X-ray flux and the soft X-ray spectra of the eruptions are reminiscent of QPEs, its mean recurrence time is as much as 22 days, and its temperature evolution during the eruptions deviates from QPEs since the pattern is colder $\rightarrow$ warmer $\rightarrow$ warmer. This source is, however, very interesting since it exhibited a transient radio emission at the beginning of one of the X-ray eruptions and repeatedly showed an indication for the X-ray absorption feature, potentially hinting at a nuclear outflow.

QPEs have been intensively monitored thanks to the space-borne X-ray observatories, such as \textit{NICER}, \textit{Chandra}, \textit{XMM-Newton}, \textit{Swift}/XRT, and \textit{eROSITA}. Single observational campaigns of the QPEs typically take only a few days, detecting only a handful of outbursts during one epoch. One of the first multi--epoch observations of the QPE source GSN 069 over 2 years showed an intriguing evolution: a decline in the count rate of the outburst phase as well as the quiescent phase \citep{Miniutti_2023}. Other long-term observations followed: eRO-QPE1 showed a relative decline by a factor of 4--10 in peak count rates over 3 years \citep{Pasham_2024} and eRO-QPE3 in a similar way. Interestingly, the monitoring of eRO-QPE2 \citep{alivestronglykickingstable} showed a rather stable level of the count rate and the recurrence time. So far, QPEs with significant periodicities have been detected in the soft X-ray domain. In the future, high-cadence monitoring in the UV domain will have the capacity to expand the parameter space towards either lower SMBH masses for the fixed periodicity and/or perturbations taking place at larger accretion disk radii for both the fixed periodicity and the SMBH mass \citep{2021arXiv211115608K,2024ApJ...963L...1L,2024ApJ...964...74S,2024SSRv..220...11W,2024SSRv..220...29Z,2024SSRv..220...24K,2025arXiv250119365Z}. This is given by the radial temperature profile of standard accretion disks, $T_{\rm d}\propto r^{-3/4}$, and by the simple scaling provided by Eq.~\eqref{eq_radius_QPE_periodicity_mass}. We refer to \citet{2024ApJ...963L...1L} for a detailed analysis. Recently, \citet{Guo_2026} showed that for the ``Ansky'' source, its UV emission is significantly correlated with the X-ray QPEs and it lags behind them by $\sim 1$ day. This strengthens the potential for the detection of new UV QPEs in the future. 

Several theoretical models were devised in an attempt to explain the physical mechanisms behind QPEs. We
can roughly divide them into two main categories:

\noindent 1) Accretion disk thermal and viscous instabilities in the region between the inner advection-dominated accretion flow (ADAF) and the outer thin disk \citep{Raj_2021,Pan_2022,2020A&A...641A.167S,2023A&A...672A..19S}.

\noindent 2) Interactions of orbiting bodies (e.g., stars, stellar-mass black holes) with the circumnuclear medium, especially the accretion disk around the SMBH. The mechanism of QPEs triggered by star-disk interactions was proposed by \cite{Xian2021}. The star-disk interactions were further analyzed in several other works, \citep[e.g.][]{Frenchie, Sukova_2021,2026arXiv260202656J}, while others focused on the mass transfer from the Roche lobe of the star \citep{2020MNRAS.493L.120K,Krolik_2022,Lu.et.al} onto the SMBH.

The latter class of models consists of a binary system with a large mass difference between the components: extreme mass-ratio inspiral (EMRI), which is applicable to a star orbiting the SMBH (the EMRI is also frequently used in the context of gravitational waves, which is not the focus of this paper).

As we point out above, long-term observations of some QPE sources imply that the accretion disk could be the result of a tidally disrupted star by the SMBH \citep{Shu_2018}. This is supported by a handful of arguments, with the key notion that a temporal decline is evident in quiescent luminosities as well as in peak luminosities. Other arguments for the stellar origin of the disk include:

\begin{itemize}
    \item high C/N (carbon/nitrogen) ratio in UV \citep{Sheng_2021} and X-ray \citep{chakraborty_2025} spectra typical of TDEs;
    \item compactness (smaller radial extent) and viscous spreading of the outer radius of the accretion disk \citep{Wevers_2025};
    \item properties of the host galaxies: both QPEs and TDEs tend to inhabit post-starburst galaxies \citep{2025ApJGilbert}.
\end{itemize}

In this paper, we revisit the association of QPEs with TDEs. Specifically, we look at the relation between the post-TDE disk formation and the rate of decrease in the QPE peak luminosities with time. In general, the peak luminosities are expected to be proportional to the instantaneous accretion rate, implying that for the typical decay in the post-TDE accretion rate, $\propto t^{-p}$ \citep{TDEs, Jankovic_2023}, where $0\lesssim p \lesssim 2$, we should see a decay in the QPE amplitudes over time as well.

This paper is structured as follows. In Section~\ref{sec2}, we briefly introduce the EMRI model. Section~\ref{sec3} presents the results of the analysis and the constraints on the parameters of the EMRI model for different QPE sources. We discuss the implications of our findings in Section~\ref{sect_discussion} and provide a summary of the main conclusions in Section~\ref{sec4}.

\section{Model setup}
\label{sec2}
\indent

Overall, QPEs are quasi-periodic, meaning that something perturbs their timing properties and their amplitudes as well. In this work, we adopted the periodic motion of a star close to the SMBH, whose dynamic is generally described using the EMRI setup. Fluctuations in the periodicity of star-disk encounters are the result of the overall star-disk geometry and the types of precession motion \citep[for a description of disk and stellar orbital precessions, see][]{Frenchie}.

\begin{figure}
    \centering
    \includegraphics[width=\linewidth]{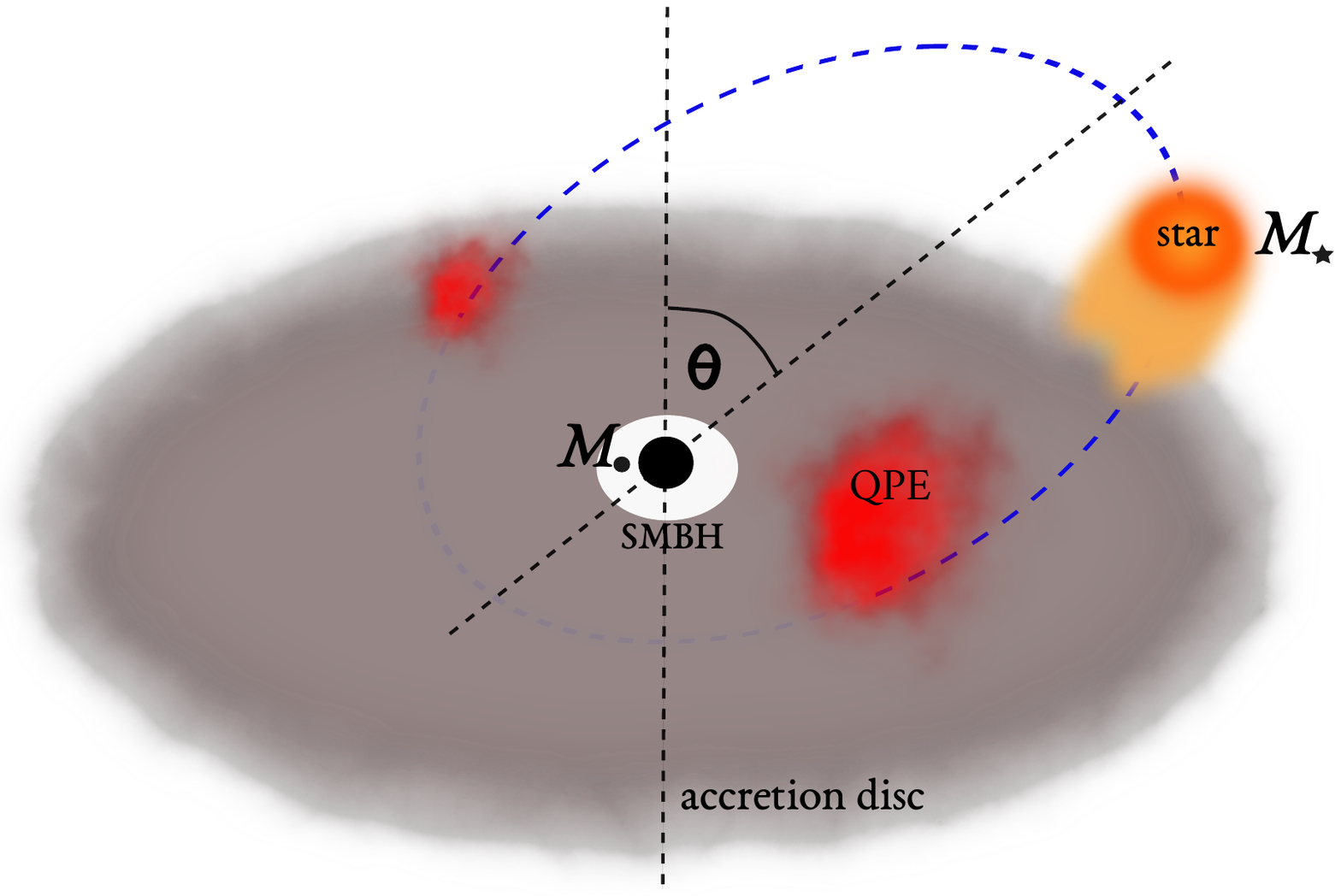}
    \caption{Scheme of an EMRI system. A body of mass $M_{\star}\ll M_{\bullet}$ follows an inclined, elliptical trajectory, on which it intersects the accretion disk twice per orbit (generally at two different radii, shown with red plumes). These collisions ablate the stellar atmosphere and push the gaseous material out of the disk plane.}
    \label{Sketch}
\end{figure}

We adopted the description of our theoretical EMRI system primarily from \cite{emritde}. The scheme of the system is illustrated in Fig. \ref{Sketch}. A solar-type star $M_\star$ with a mass of $M_{\star} = M_1\,\text{M}_\odot$ orbits the SMBH with mass $M_\bullet = 10^6 M_{\bullet,6}\,\text{M}_\odot$. Here, the dimensionless quantity $M_1 = 10$ describes a star of ten solar masses, similarly with $M_{\bullet,6}$, where $M_{\bullet,6} = 10$ means $10^7\,\rm M_\odot$). The star has an orbital velocity in units of the percents of light speed; specifically, for a nearly circular orbit, we have $v/c \simeq 0.1 (a/100\,\rm R_{\rm g})^{-1/2}$. The star regularly crashes with an accretion disk twice per orbit with a period of $P_{\rm orb}=2\,P_{\rm QPE} \sim \pi\,(GM_{\bullet}/c^3)(a/\rm R_{\rm g})^{3/2}\sim 4.3\, M_{\bullet,6} (a/100\,\rm R_{\rm g})^{3/2}$\,h and ejects a cloud of gas that expands adiabatically \citep{emritde}. Thermal emission of the expanding cloud peaks in the soft X-ray domain (see also Appendix~\ref{apendixb} for example spectra).

The disk is assumed to be geometrically thin and optically thick with the vertical scale-height $h$ that follows from \citet{accpower}, we have
\begin{equation}
\label{scaleheight}
    \frac{h}{r} \approx 2.3 \times 10^{-2}\, \dot{m}_{-1} \left(\frac{r}{100\,\rm R_\text{g}} \right)^{-1},
\end{equation}
where $r$ is the radial distance from the SMBH. This formula is approximately valid for the accretion rate of $\dot{m} \in (10^{-3},\,1)\,\dot{m}_\text{edd}$ expressed relative to the Eddington accretion rate. Here, Eq.~\eqref{scaleheight} is valid for a stationary extended accretion disk. In particular, it neglects the effect of star–disk impact near the tidal radius, where the returning debris also rejoins the disk, so the disk there is expected to be thicker than in the stationary case due to the debris-disk heating. We often use the scaling relation of accretion rate compared to 0.1 of the Eddington value, $\dot{m}_{-1} = \dot{m}/(0.1\,\dot{m}_\text{edd})$. The Eddington accretion rate comes from the Eddington luminosity,
\begin{equation}
    \dot{m}_\text{edd} = L_\text{edd}/(0.1\,c^2) \approx 0.02\,M_{\bullet,6}\,{\rm M_{\odot}\,yr^{-1}}.
\end{equation}

We also assumed that a transient accretion disk is created from the debris of another main sequence (MS) star with the mass $M_{\star,\text{dis}} = M_\text{dis}\,\rm M_\odot$ that was tidally disrupted beforehand. It probably arrived on a parabolic orbit sufficiently close to the SMBH. The magnitude of the tidal forces is directly correlated to the impact parameter, $\beta$, defined as the ratio of the tidal radius to the pericenter distance, $\beta = r_{\rm t}/r_{\rm p}$. For the sake of simplicity, we state  that the condition for complete destruction of an MS star is $\beta \gtrsim 1$ \cite[for the discussion about the critical $\beta$ for a complete TDE, see][]{Rossi_2021}. The tidal radius, $r_{\rm t}$, is the distance from the SMBH, where the gravitational pull of the SMBH is comparable to the self-gravity of the star \citep{TDEs}, expressed as
\begin{equation}
    r_{\rm t} \simeq R_{\star,\text{dis}} \left( \frac{M_\bullet}{M_{\star,\text{dis}}} \right)^{1/3},
    \label{beta}
\end{equation}
where $R_{\star,\text{dis}} = R_\text{dis}\,\rm R_\odot$ is the radius of $M_{\star,\text{dis}}$. After the circularization of the debris, the accretion rate through the disk is set by the interplay between the ongoing fallback of the debris, the viscous evolution of the disk, and the potential onset of disk outflows. Therefore, it does not need to track the canonical debris fallback rate, $\propto t^{-5/3}$, leading to  $\dot{m} \propto t^{-1}$ at later times \citep[e.g.,][]{Balbus_2018, TDEs, Shen_2014}. Thus, we modeled the accretion rate as $\dot{m} = \dot{m}_0 (t/t_{\rm fb} )^{-p}$, where  $\dot{m}_0$ is the initial accretion rate and $t_{\rm fb}$ is the characteristic fallback timescale defined as the orbital period of the most bound debris,

\begin{equation}
t_{\rm fb} \sim 56\,M_{\bullet,6}^{1/2}  M_{\rm dis}^{1/5}\,\text{days}\,.  
\end{equation}

We treated $p$ as an effective decay index in the observation window and explored a range of values $p \in (1/2, 9/4)$, spanning viscous-dominated evolution and fallback-dominated limits. W also note that partial disruptions can yield steeper fallback decays \citep[see e.g.,][]{Balbus_2018,Coughlin_2019, TDEs}. We separately accounted for the time-dependence of the disk mass distribution due to viscous spreading, described in Section~\ref{decline_section} and expressed via Eq.~\eqref{eq_sigma_decrease}. 

Since we are varying the parameters of the EMRI system in our calculations throughout this paper, we present the range of values in Table \ref{fiducial} for a better overview.

\begin{table}[htp]
    \caption{\centering Quantities with their fiducial values denoted with (f).}
    \begin{tabular}{c|c|c}
        \hline \hline
        Quantity & Notation & Value\\
        \hline
        SMBH mass [$\rm M_\odot$] & $\log{M_{\bullet}}$ & 5, 6(f), 7 \\
        Stellar mass [$\rm M_\odot$] & $M_\star$ & 0.3, 1(f), 8 \\
        Disrupted stellar mass [$\rm M_\odot$] & $M_\text{dis}$ & 1(f), 5 \\ 
        Tidal radius [$\rm R_\text{g}$] & $r_\text{t}$ & 6\text{ - }600, 60(f)\\
        Viscosity parameter & $\alpha$ & 0.1(f) \\     
        \hline      
    \end{tabular}
      \label{fiducial}
\end{table}
Here, we consider the behavior $\dot{m}$ after it reaches its power-law state. This will happen after the circularization timescale, which we approximate as $T_\text{cir} \approx 10\,\rm t_\text{fb}$ \citep{TDEs}. The initial accretion rate  \citep[see][]{emritde} can be expressed as
\begin{equation}
    \dot{m}_0 \sim 80\,M_{\bullet,6}^{-3/2} M_\text{dis}^{4/5}\,\dot{\rm m}_\text{edd}.
\end{equation}

In general, the orbiting star ($M_1$) on a quasi-circular orbit comparable to the radius of the TDE disk will experience a gravitational wave inspiral on a timescale of $10^5\text{--}10^6\,$yr \citep{1964PhDT........51P}. This is much longer than the timescales of  interest here, on which the QPE amplitudes change and later diminish entirely. On the other hand, the average interval between consecutive TDEs in a galactic nucleus is $10^4\text{--}10^5$ years \citep{Magorrian1999}. This brings us to the conclusion that most EMRIs will probably experience a TDE before getting too close to the SMBH to undergo a Roche-lobe overflow onto the SMBH and eventually end up being destroyed by its gravitational field. In the following sections, we consider the evolution of the EMRI system on timescales of years.

For an analysis of the behavior of QPEs, we adopted the EMRI model in post-TDE hosts and utilized a time-dependent description of the system parameters.

Following \citet{emritde} and using the relation,
\begin{align}
    L_\text{char} \approx  \frac{L_\text{edd}}{3} \frac{(R_{\star}^2 h)^{1/3}}{a},
\end{align} 
the characteristic luminosity of QPEs can be expressed as
\begin{align}
    L_\text{char} \sim \overbrace{1.4 \times 10^{41}\,R_1^{2/3} M_{\bullet,6} P_\text{QPE,d}^{-2/3}} ^{C}\dot{m}_{-1}^{1/3}\,\text{erg\,s}^{-1},
    \label{Lqpe}
\end{align}
where we use $C$ to mark the amplitude of $L_\text{char}$ that does not change significantly over time; $R_1$ is the radius of the star; and $P_\text{orb,d} = P_\text{orb}/(24\,\text{h})$ is the period of the star. The total emitted energy during one disk transition is estimated as
\begin{equation}
    E_\text{ej} \sim 8.2 \times 10^{45}\frac{R_1^2 P_\text{orb,d}^{1/3}}{\alpha_{-1} \dot{m}_{-1} M_{\bullet,6}^{1/3}}\,\text{erg}\,,
    \label{E_ej}
\end{equation}
where the viscosity parameter is $\alpha_{-1} = \alpha/0.1$. Both of these expressions depend on $\dot{m}$, which is generally a function of time, $\dot{m} = \dot{m}(t)$.

In the following section, we investigate how the amplitudes of the eruptions (i.e., the peak luminosities) decline over time based on the decline in $\dot{m}$ as well as the star losing its mass due to ablation. In addition, we also describe how we designed the scheme  to construct simple artificial light curves from the characteristic luminosities and timing properties. 

\section{Results}
\label{sec3}

In the following subsections, we aim to quantify the consequences of a time-dependent accretion disk solution on the eruption amplitudes. First, the circularized disk needs to have a certain radial extent to ensure that EMRI crashes with it (as discussed in Sect.~\ref{constraints}). Subsequently, we give examples of simple artificial light curves that are in agreement with the GSN 069 amplitude and the ratio of the decline and rise times (Sect.~\ref{modeling of LCs}). Here, the main working idea is that the accretion rate decreases with time (as discussed in Sect.~\ref{decline_section}) and how this affects emerging bolometric and X-ray luminosities (Sect.~\ref{long-term-x-ray}). The geometry of the system is also an important parameter in the emission mechanism that is closely related to the precession mechanisms (Sect.~\ref{geometry}). Finally, we discuss the EMRI being stripped of its material, which impacts the luminosities of the gas clouds since $M_1$ gets smaller over time (see details in Sect.~\ref{mass loss}).

\subsection{Constraints}
\label{constraints}
\indent

We estimated the semimajor axis $a$ of $M_1$ based on the SMBH mass and the period, which is taken as twice the median QPE recurrence time; $P_\text{orb} = 2\langle P_\text{QPE}\rangle$. In all cases, we considered nearly circular orbits, so the eccentricity would be $e \sim 0$. As the secondary impacts the accretion disk, its pericenter has to be smaller than the outer radius of the disk \cite[for a discussion about the EMRI orbit parameters, see][]{Karas_2001, Syer1991}. The expression for the outer radius, $R_\text{out}$, can be estimated according to \citet{LT_prec} and written on the right side of the equation,
\begin{align}
     R_\text{p} = a(1 - e) < \underbrace{94\,M_{\bullet,6}^{-2/3} M_\text{dis}^{7/15} \beta^{-1} \Bigl(\frac{t}{t_\text{vis}}\Bigr)^{2n-2}}_{R_\text{out}},
\end{align}
implying 
\begin{align} 
     \frac{5}{2}P_\text{orb,d}^{2/3} (1 - e)M_{\bullet,6}M_\text{dis}^{-7/15} \beta \Bigl(\frac{t}{t_\text{vis}}\Bigr)^{-3/5} < 1\,,
     \label{condition}
\end{align}
which is the basic collision condition and where $t_\text{vis}$ is the viscous timescale related to the hydrodynamics of the disk evolution (see Sect. \ref{decline_section}) and $n \sim 1.3$.
For a more detailed discussion covering the conditions of an EMRI crossing the post-TDE disk and the extent of $R_\text{out}$, we refer to \citet{mummery2025}. This condition is well suited for eRO-QPE2 and RXJ1301.9+2747, even for early times, and is valid for eRO-QPE1 and GSN 069 if we allow greater $M_\text{dis}$. The same is valid for the systems eRO-QPE3,4 and Ansky. However, AT2019qiz and Swift J0230+28 strongly violate our condition. In either case, the EMRI can only crash into the disk after its outer radius has spread far enough outward, given the evolution of the surface density $\Sigma(t)$. Regarding the dependence on time, $t$, the collisions can occur after the gaseous material has been circularized into the disk. This happens after $T_\text{cir}$. A relevant circularization timescale has been introduced in the work of, for instance, \citet{Syer1991}. In fact, during the initial phase of circularization, only a single transit can occur per each orbit \citep{Vokrouhlicky1998}.

\subsection{Modeling of the light curves}
\label{modeling of LCs}

We created an asymmetric Gaussian-like light curve profile, so that the ratio of the decline and rise time is $\approx 1.3$, as estimated in Appendix \ref{apendixa} for a particular QPE (although we note that similar values were estimated for eRO-QPE2 and GSN 069). The analytical equation of the profile follows from \citet{barlow2004asymmetric}. The value of $L_\text{char}$ follows from Eq.~\eqref{Lqpe}.

\begin{figure}
    \centering
    \includegraphics[width=\columnwidth]{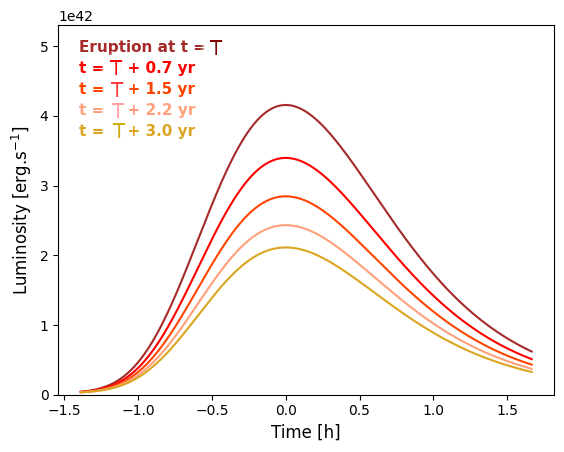}
    \caption{Surrogate light curve evolution during 3 years at equidistant time intervals modeled using a quasi-Gaussian profile with a given duration and an amplitude given by Eq.~\eqref{Lqpe}.}
    \label{Artificial light curves}
\end{figure}

Figure~\ref{Artificial light curves} shows the theoretical light curve profiles for GSN 069-like system parameters ($M_{\bullet} \sim 10^{6}\,\rm M_\odot$, $t_\text{dur} \sim 1.3\,$h, $\dot{m} \sim 0.1\,\dot{m}_\text{edd}$) for the temporal separation of 3 years such that their relative amplitude change is $\approx 2$. After integrating the light curves, the escaped energy, $E_\text{ej}$, is almost the same as the integrated energy at $t = T$, but at $t = T + t_\text{delay}$ ($t_\text{delay} = 3\,$yr, see
Sect.~\ref{decline_section}), $E_\text{ej}$ is only $1/5$ of the integrated luminosity/energy fraction (i.e., transferring more energy into the kinetic energy of the cloud, etc.).

\subsection{Long-term amplitude bolometric decline}
\label{decline_section}

Several QPE sources were observed in multiple epochs separated by months and years (\citealt{Miniutti_2023, alivestronglykickingstable}). On timescales of years, a visible decline in the peak count rates and the quiescent emission from the disk was measured in many sources. This is one of the strong arguments supporting the idea that QPEs could be tied to post-TDE accretion disks since the magnitude of the disk accretion rate, together with the power of the eruptions, gets weakened.

\begin{table}[htp]
    \caption{\centering Peak amplitude declines in QPEs.}
    \centering
    \resizebox{\columnwidth}{!}{
    \begin{tabular}{c|c|c|c} 
    \hline
    \hline
        Source & Relative decline$^a$ & Time span$^b$ & \# Epochs \\
        \hline
        eRO--QPE1 & 4-10x   & 3 yr & 6 \\ 
        eRO--QPE2 & 1x   & 3.5 yr & 5 \\ 
        GSN 069   & 1.5x  & 2 yr & 2 \\ 
        eRO--QPE3 & 6x    & 2 yr & 5 \\
        Ansky     & 4x    & 1 yr   & 2 \\ \hline
    \end{tabular} 
    }
    \tablefoot{\\
    \tablefoottext{a}{Ratio of count rates in the first and last epochs of observation. Second, eRO-QPE2, did not show any conclusive decline.} \\
    \tablefoottext{b}{Amount of time between first epoch and last epoch. The papers describing the relative declines are: eRO-QPE1 and eRO-QPE2 \citep{Pasham_2024}, GSN 069 \citep{Miniutti_2023}, eRO-QPE3 \citep{arcodia2024}, and Ansky \citep{2025ApJZhu}.}
    }
    \label{table}
\end{table}

\begin{figure*}[htp]
    \sidecaption
    \centering
    \includegraphics[width=0.65\linewidth]{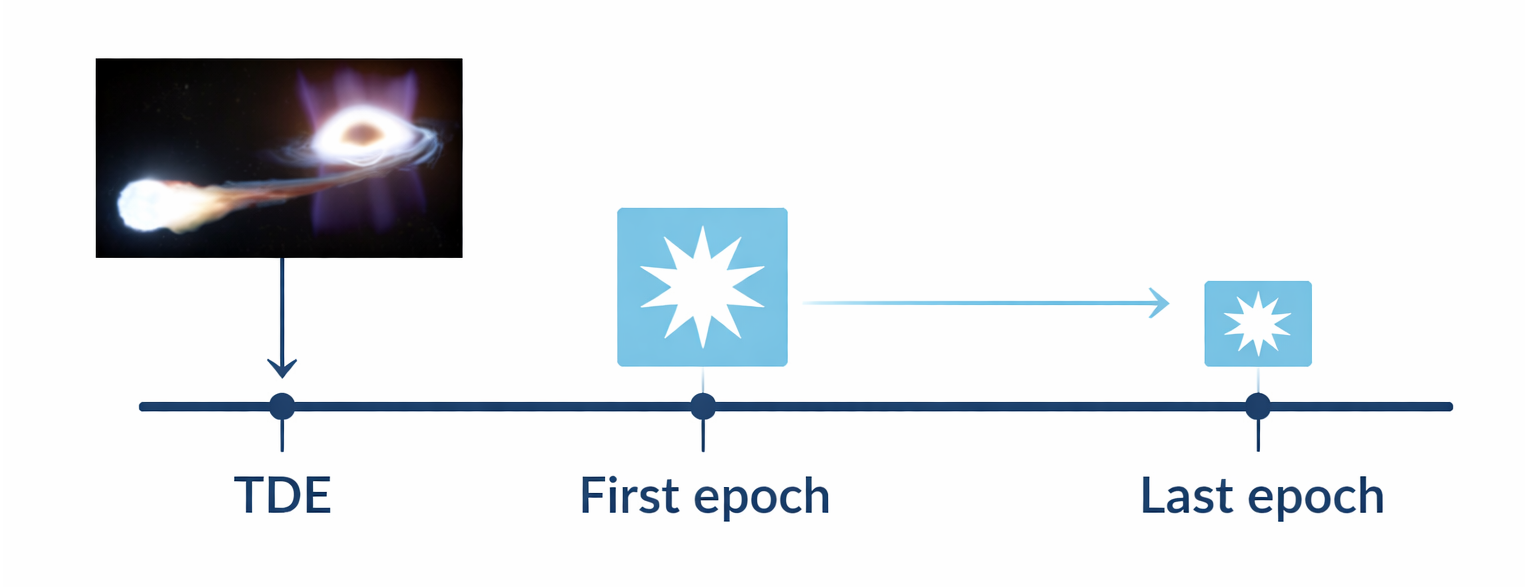}
        \caption{Timeline of the QPE system evolution. The first interval with the period, $T$, is the variable, second interval is chosen to be $t_\text{delay} = 3\,$yr. Credit: STScI.}
    \label{timeline}    
\end{figure*}

Taking this feature into account, the amplitude drop depends on two intervals. One takes place some time after the disruption ($t = 0$) of $M_{\star,\text{dis}}$ when the eruptions would be monitored for the first time (at time $t = T$). The second interval covers all the theoretically observed epochs of the QPE system and ends with the last ($t = T + t_\text{delay}$). As we can see in Table~\ref{table}, the coverage of a typical long-term observation spans 3 years. We kept the second interval constant, namely, $t_\text{delay} = 3\,$yr. A simple timeline is shown in Fig.~\ref{timeline}.

\begin{figure}
 \centering
    \includegraphics[width=\columnwidth, center]{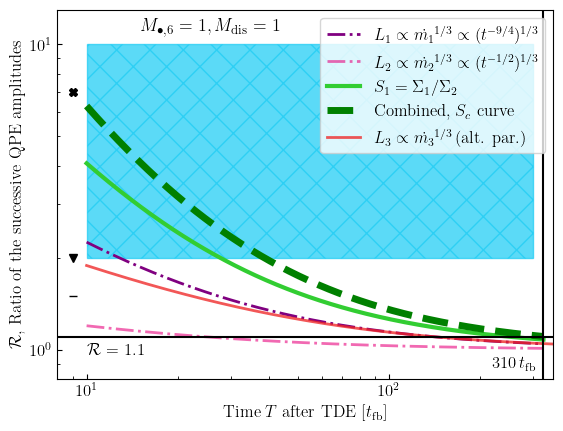}
    \caption{Theoretical ratio of the peak luminosities $\mathcal{R}(T)$. A thin solid red line corresponds to alternative parameters: $M_{\bullet,6} = 0.1, M_\text{dis} = 10$, the remaining lines correspond to $M_{\bullet,6} = 1,\, M_\text{dis} = 1$. Luminosities of $L_1-L_3$ are power-law functions of $\dot{m}$, while the accretion rate is expected to drop by $\dot{m} \propto t^{-p}$, where $p \in \{9/4, 1/2, 6/5\}$ and the corresponding accretion rates are $\dot{m}_1$, $\dot{m}_2$, and $\dot{m}_3$, respectively. The solid green thick line shows the decline of the surface density of the disk due to viscous spreading, $\Sigma$, and the thick dashed line represents the combined effect of the $\Sigma$ decline and canonical ($-1.2$) $\dot{m}$ decline that crosses the value of 2 at $40\,\rm t_\text{fb}$. The blue rectangle covers values 2-10. The minimum value of $T$ is at $10\,\rm t_\text{fb}$. The value of $\mathcal{R} = 1.1$ corresponds to our minimum expected detectable ratio and crosses the $S_\text{c}$ line at 310$\,\rm t_\text{fb}$. The markers on the left side show the decline values for individual sources listed in the second column of Table~\ref{table}. Finally, x,  v, - belong to eRO-QPE1, eRO-QPE3, and GSN 069, respectively.}
    \label{Amplitude ratio}
\end{figure}
Using the approximation of $L_\text{char}$ from Eq.~\eqref{Lqpe},
we can plot the drop in these values with time based on how long after the disruption event, $t=T$, the first epoch of light curves could be observed. In Fig.~\ref{Amplitude ratio}, we can see the result in terms of $\rm t_\text{fb}$.

We state that if the ratio of peak luminosities is less than 1.1 (10$\%$ change), the decline would be observationally indiscernible, resulting in stable luminosities over time. In this case, after 310$\,\rm t_\text{fb} \sim 50\,$yr ($M_{\bullet,6} = 1, M_\text{dis} = 1$), there would be no measurable decrease in luminosities for the curve, $S_{\rm c}$ (explained below).

The luminosity amplitude depends on two power-laws: $\dot{m}$ changing with the power-law index, $p$, and the luminosity itself depending on some power-law of $\dot{m}$. Covering the extremes of the parameter space of $p \in (1/2,\,9/4)$ shows that for any of its values, the ratio of theoretical typical luminosities is small (curves $L_1$, $L_2$, $L_3$ in Fig. \ref{Amplitude ratio}). This ratio is called $\mathcal{{R}}$ and we define it using the successive luminosities as

\begin{equation}
    \mathcal{{R}}(T) = \frac{L(T)}{L(T+t_\text{delay})}
.\end{equation}
The crossing of the light blue rectangle (Fig.~\ref{Amplitude ratio}) with the borders of $\mathcal{{R}} = 2\text{--}10$ does not happen anywhere for $L_2$. For $L_1$, a two-fold drop would be possible given the first epoch of observation had started $15\,\rm t_\text{fb}$ after the disruption, thereby requiring a quite small observing interval. Lastly, we plotted the $L_3$ curve that has the same $L-\dot{m}$ power-law relationship, but we upscaled the disrupted star and downscaled the SMBH by one order of magnitude each. The result is similar to what was found for $L_1$. However, there is one caveat: as the disk is viscously spreading and the star plunges into the disk approximately at the same distance from SMBH, the place of impact will encounter smaller surface density every time, affecting the emerging outburst luminosities.

In light of recent hydrodynamical solutions from \citet{guolo2025}, a time-dependent post-TDE accretion disk parameterization can be applied in this case. Useful time-dependent relations of $\Sigma,\,L_\text{char}$ and other relevant disk properties have been quantified. As $L_\text{char} \propto E_\text{ej} \propto M_\text{ej} \propto \Sigma$ \cite[][energy of the ejecta]{emritde}, we can verify the long-term decline based on the inferred scaling relations, since the results from the simulations predict
\begin{equation}
    \Sigma(r=\text{const},t) \propto (t/t_\text{vis})^{-n},
    \label{eq_sigma_decrease}
\end{equation}
where $t_\text{vis}$ is the viscous timescale and $n = 1.3$. The value of $t_\text{vis}$ can vary over multiple orders of magnitude and for GSN 069-like system parameters, it is estimated to be $t_\text{vis,gsn} \sim 2300 \pm 300\,$d. With this in mind, we separated the density effect and show its decline in Fig.~\ref{Amplitude ratio} (i.e., we want to see how the luminosity is affected by the $\dot{m}$  and $\Sigma$ decreases with time separately). We chose the value of $t_\text{vis}$ comparable to $T_\text{cir}$. A clear cross of the rectangle appears for $\mathcal{{R}} \in (2, 4)$ with the $S_1$ (thicker green) curve. Combining both effects together gives us the thickest curve ($S_{\rm c}$), reaching the largest values of $R$ (up to $\sim 7$). This shows that including the secular decrease of $\Sigma$, leads to a faster evolution of the luminosity amplitude than in the cases driven solely by the decrease in $\dot{m}$, since $\dot{m}$ and $\Sigma$ both decrease with time, further reducing $L_{\rm char}$. While this does not explain the peak amplitude drop value for eRO-QPE1, accounting for additional effects in this simplified picture can yield higher values of $\mathcal{R}$ (EMRI inclination, eccentricity, etc.). The values for other QPEs can be easily reached for a wide range of parameters with $M_\bullet$ and $M_{\rm \star,dis}$. Lastly, it is illustrative to show the basic disk properties across the whole range of $T,$ so we can get an idea of the regime the disk is currently in. These results are given in Table~\ref{properties} based on Eq.~\eqref{scaleheight}.

\begin{table}[htp]
    \centering
    \caption{\centering Accretion rate and scale-height of the post-TDE accretion disk at five different epochs.}
\resizebox{0.6\columnwidth}{!}{
\begin{tabular}{c|c|c}
\hline 
\hline
     T [$\rm t_{\rm fb}$] & $\dot{m}$ [$\dot{m}_{\rm edd}$] & $h(100\,\rm R_{\rm g})$ [$R_\odot$]  \\ \hline
     10 & 1.7 & 80 \\
     80 & 0.1 & 2.6  \\
     150 & 0.02 & 0.9  \\
     220 & 0.01 & 0.5  \\
     290 & 0.006 & 0.3  \\ \hline
    \end{tabular}
    }
    \label{properties}
\end{table}

\subsection{Long-term amplitude X-ray decline}
\label{long-term-x-ray}

\begin{figure}
    \centering
    \includegraphics[width=\columnwidth]{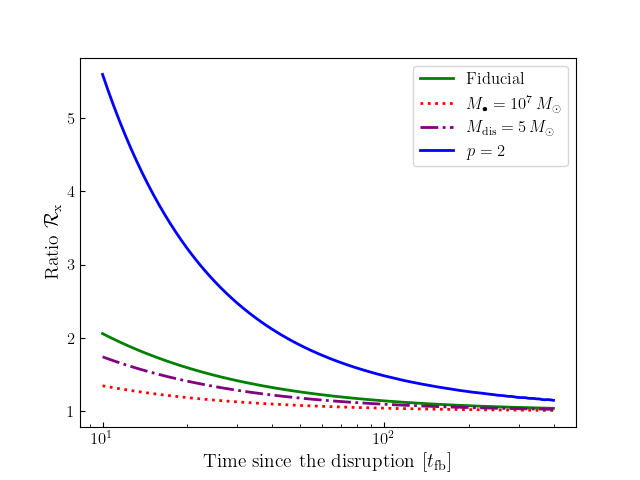}
    \caption{Analogously to Fig.~\ref{Amplitude ratio}, a long-term X-ray amplitude decline is shown here for the monitoring delay of 3 years. The fiducial parameters are as follows: $M_{\bullet} = 10^6\,\rm M_\odot$, $M_\star = 1\,\rm M_\odot$, $P_\text{QPE} = 24\,$h, $M_{\text{dis},\star} = 1\,\rm M_\odot$, and $p = 1$, $\alpha = 0.1$. See details in Appendix~\ref{apendixb}.}
    \label{xratio}
\end{figure}

A plausibly better comparison of the theoretically calculated QPE long-term amplitude decline with the observational data could be obtained by expressing QPE amplitudes in the soft X-ray domain. To determine these luminosities, we calculated the standard gas-pressure dominated accretion disk properties (see Appendix \ref{apendixb}), post-shock temperature of the ejected gas cloud, and the estimated radius of the cloud as $R_\text{cl} \sim 10^{11}\,$cm, which agrees with the observations \citep{Frenchie}. We keep $R_\text{cl}$ constant during the outburst peak for simplicity. According to the standard blackbody spectral energy distribution, we obtained $(0.2, 2)\,$keV luminosities. We then compared these results to their 3-year delayed counterparts. The result can be seen in Fig.~\ref{xratio}.

To reproduce the largest amplitude drops in 3 years, changing the mass of the SMBH does not yield big enough changes in $\mathcal{R}_{\rm x}$. We do not explicitly show the effect of viscous spreading on the amplitudes here since we would only replicate the $S_1$ curve from the Fig.~\ref{Amplitude ratio}. However, the accretion rate power-law index, $p$, has the leading-order effect again. In combination with the decreasing surface density due to the viscous spreading of the disk, the largest values of $\mathcal{R}_\text{x}$ can be reproduced (see also Fig.~\ref{Amplitude ratio}). The value of $\mathcal{R}_\text{x}$ is analogically defined as the ratio, $\mathcal{R}$, given in Section~\ref{decline_section}, but using the soft X-ray luminosity instead of the bolometric luminosity. However, the bolometric-based conclusions remain robust since the ratio $L_{\rm x}/L_{\rm bol}$ evolves only mildly across the relevant parameter space (see also Appendix~\ref{apendixb}).

\subsection{Inclination of the EMRI}
\label{geometry}

QPE models mostly consider collisions inclined perpendicularly ($90^{\circ}$) with respect to the disk. In this subsection, we briefly discuss the simplest implications of the orbit inclination. From Eq.~\eqref{Lqpe}, we notice that $L_{\text{char}} \propto v_{\text{rel}}^{-2}$, meaning the inclination of the EMRI towards the disk would change the relative velocity. Noticing that the ratio of $v_\text{rel,90}$ and $v_{\text{rel},\theta}$ (perpendicular component) is inclined at an angle $\theta$ (see Fig.~\ref{Sketch}) between the axis of the disk, the normal to the orbital plane is
\begin{equation}
    \frac{v_{\text{rel},\theta}}{v_\text{rel,90}} = \sqrt{1 - \cos\theta}\,.
\end{equation}

For angles close to $90^{\circ}$, the change is negligible; however, as we approach small angles the ratio can become tiny; in particular, changes in the inclination angles of an already inclined EMRI can cause significant variations in luminosities. However, we have to be cautious about small angles since the scale-height of the disk constrains the orbit of our star. With a small enough $\theta$, we would have a coplanar orbit of the immersed star and this QPE mechanism would not apply here (i.e., no collisions could occur).

The precession in the system has an impact on the luminosity as well. This changes the position of the impact with the disk and the relative velocity of EMRI. The farthest and closest places of the impact from the SMBH are the apo-, $R_\text{a} = a(1+e)$, and pericenter, with their ratio being $(1+e)/(1-e)$. As $L_\text{char} \sim 1/a$, the luminosity ratio of the impact places at $R_\text{p}$ and $R_a$ are $\approx 1$ for quasi-circular orbits, but moderately elliptical orbits affected by precessions could cause variability of burst amplitudes by larger factors.

For a more detailed discussion about the EMRI inclination towards the disk, we refer to Section 2 of \citet{suzuguchi2025}. In that work, the evolution of the QPE duration, luminosity, and temperature is demonstrated for a disk created after a TDE and for a large parameter space of $\theta$. However, their investigations concern only Eddington and super-Eddington disks.

\subsection{Mass loss from an ablation of the star}
\label{mass loss}
\indent

As the star crosses the disk supersonically, the ram pressure starts to strip the outer layers, stealing the material. The amount of stripped material per orbit can be similar to the accretion rate of the disk. The mass and energy of the stellar material mixing with the disk can either enhance or stabilize the accretion \citep[see the discussion in][]{Linial_2024}. From hydrodynamic simulations showing the interaction between the star and the ejecta of a supernova explosion \citep{Liu2015}, some QPE models (e.g., \citealp{Frenchie}) use derivations of this approach to estimate the mass loss in the star. The ablated stellar material is proportional to the ram pressure and depends on the binary separation, $a$ \citep[cf.][]{2025MNRAS.540.1586K}.

However, recent hydrodynamical simulations \citep{yao2024} offer interesting results for mass-loss per collision that are one order of magnitude higher than the previous ones. They perform similar calculations to those covering supernova-star interactions, but with two major differences: 1) the ram pressure of star-disk interaction is generally much lower and 2) the collisions are successive with not enough time for the star to adjust its outer layers into the local thermal equilibrium. After numerous collisions, the mass loss is almost 20 times larger than it was at the beginning, while the relative mass loss per collision stabilizes at a constant value. This has critical implications for the lifetime of a star in the QPE system that are
estimated to be as short as a few decades. The formula for mass loss derived in \cite{yao2024} is

\begin{equation}
    \frac{M_\text{debris}}{M_1} \approx \frac{0.06}{f_\text{M} \eta} \frac{E_\text{ej}R_1}{G M_1^2} \Big[ 1 + \frac{h}{R_1} \Big]^{-1}.
\end{equation}

The product $f_\text{M}\eta$ contains the scaled size of the debris and radiative efficiency and is approximated as $\sim 1$. Distinct from the more general approach towards the values of $\eta$, we leave it at the typical value of $\sim 0.1$. The fractional loss per transition depends on $E_\text{ej}$ which we can estimate with Eq. \eqref{E_ej}. We can scale this formula into the following form,

\begin{multline}
    M_\text{debris} \sim 2.4 \times 10^{-6}\,\rm M_\odot\,\rho_{\rm d, -6} (M_{\bullet,6}/P_\text{orb,d})^{2/3} M_1^{2.2} \\
    \times \Bigl(\frac{h}{R_1} \Bigr) \Bigl(1 + \frac{h}{R_1} \Bigr)^{-1},
\end{multline}
where $\rho_{\rm d, -6}$ is the disk midplane density scaled to $10^{-6}\,{\rm g\,cm}^{-3}$. The midplane density was derived in \citet{accpower}, expressed as

\begin{equation}
    \rho_{\rm d} \sim 8.7 \times 10^{-7}\,\text{g\,cm}^{-3} \frac{\dot{m}}{\alpha_{-1} P_{\rm orb,d}} \Bigl(\frac{h/R_1}{10^{-2}}  \Bigr)^{-3},
\end{equation}
where $h/r$ is the aspect ratio. It is important to note here that the accretion rate still declines as $\dot{m} \propto (t/t_{\rm fb})^{-p}$. As a result, $\rho_{\rm d,-6}$ and $h$ are functions of time, together with $M_1$, of course. For MS stars, we would have $R_1 \propto M_1^{0.8}$.

\begin{figure}
    \centering
    \includegraphics[width=\columnwidth]{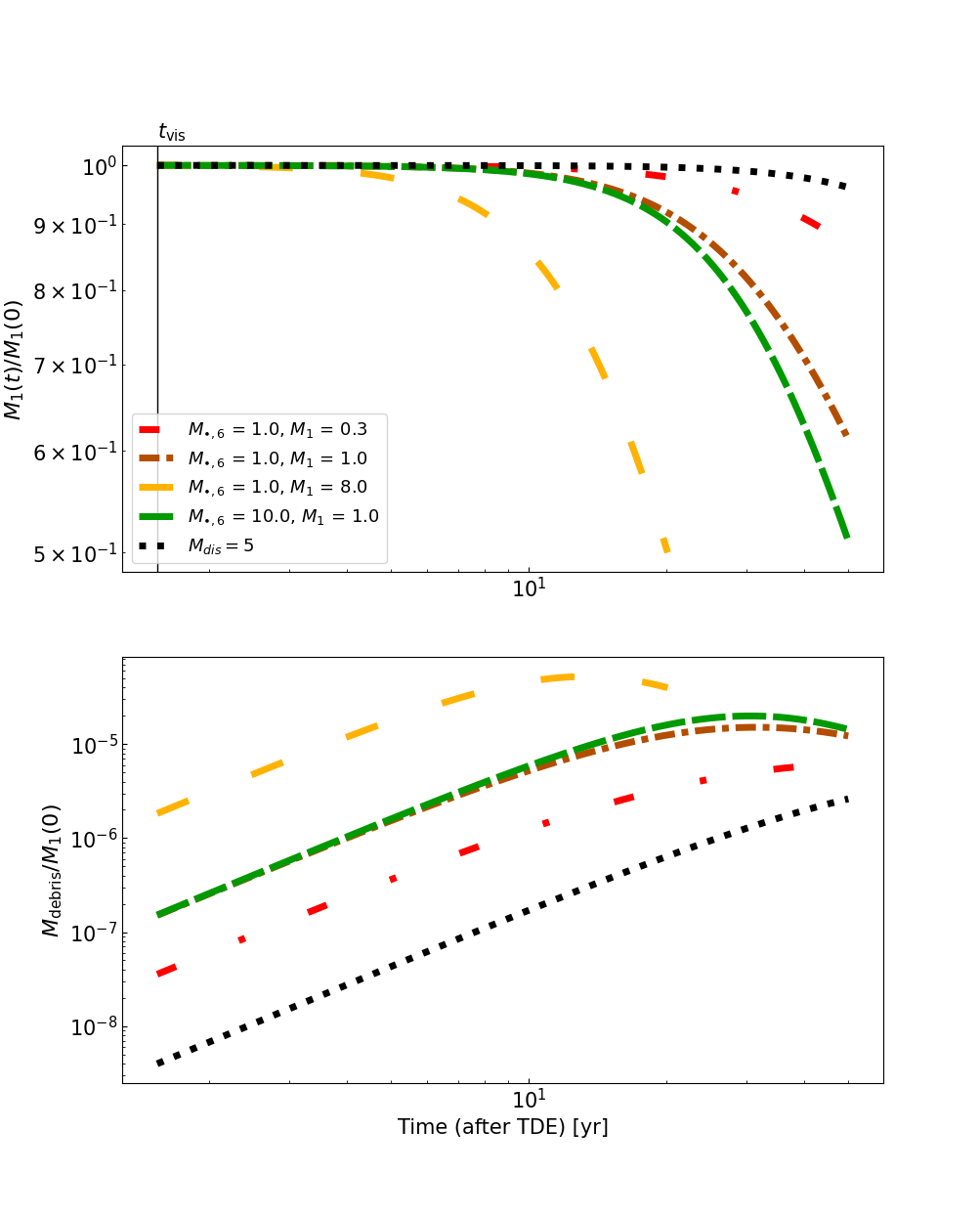}
    \caption{Ablation rate of the stellar mass after $t_\text{vis}$. \textit{Top:} Remaining stellar mass scaled to $M_{\star} = M_1(0)$ evolves with time. \textit{Bottom:} Mass of the debris lost during the disk passages, scaled to $M_{\star}$ again. The curves that reach the half on the $y$-axis (top panel) do not continue further since the star lost a significant amount of mass. We started the mass loss after $t_\text{vis}$, which is a point in time where $\rho_{\rm d}$ in a post-TDE disk starts to decay with a power-law of $n$. The first four curves are for the $M_{\rm dis} = 1,$ but the last one is for $M_{\rm dis} = 5$ and $M_{\bullet,6} = 1, M_1 = 1$.}
    \label{Mass loss}
\end{figure}

From Fig.~\ref{Mass loss}, we can infer that in just a decade, a heavy star $M_1 = 8$ is robbed of half of its mass, implying a "timescale of ablation" of two decades for heavy stars at their maximum. On the other hand, lower boundary solar-type stars would survive approximately 100 years. What is interesting is that the $M_{\bullet,6}$ has a small impact on the resulting curves, meaning that all variables in $M_\text{debris}$ dependent on SMBH mass together yield  a weak power-law in the end. For fiducial parameters, the lifetime is $\approx 40\,$yr, roughly agreeing with the estimates from \citet{Linial_2025}.

If we upscale the mass of the disrupted star $M_\text{dis}$, we reach higher longevity of the orbiting star, since $t_{\rm fb} \propto M_{\rm dis}^{1/5}$ and $\dot{m}_0 \propto M_{\rm dis}^{4/5}$ have the effect of lowering the $M_\text{debris}(t)$. As we can see, the mass loss of tens of percents can be significant on timescales lasting several years. As $L_\text{char} \propto R_1^{2/3} \propto M_1^{8/15}$, the mass loss of $M_1$ can have a noticeable impact on the long-term changes in QPE luminosities. From the $R_1$ power-law in Eq.~\eqref{Lqpe}, we see that characteristic luminosity decreases with $L_\text{char} \propto M_1^{8/15}$.

\subsection{Mass feeding to the disk}

 To this point in our study, we have considered a transient accretion disk with a declining accretion rate as a consequence of the previous tidal disruption. The changing $\dot{m}$ was considered as a primary factor in the long-term evolution of the QPE amplitudes. However, for a wide range of inclination angles of the EMRI, the majority of ablated stellar material can get mixed into the disk, enhancing the mass of the disk annulus around the places of impact. \cite{Linial_2024}  considered the accretion of material stripped from the star towards the accretion disk and found a coupled star-disk evolution that yields nearly constant accretion rates and even rising ones \citep[see also][]{Lu.et.al}.

\begin{figure}
    \centering
    \includegraphics[width=0.9\columnwidth]{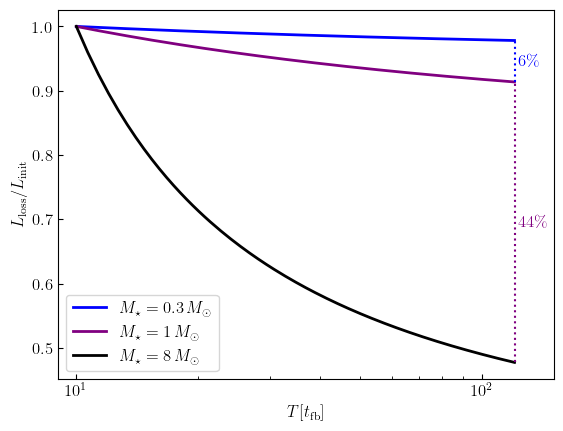}
    \caption{Relative luminosity decline of QPEs purely due to the ablation of a stellar body, i.e., a decrease in stellar mass. The lines stand for different initial stellar masses according to the legend. The accretion rate is set constant, $\dot{m} = 0.1\,\dot{m}_{\rm edd}$. The luminosity evolution with time ($L_\text{loss}$) is scaled to the initial luminosity ($L_\text{init}$).}
    \label{mdotconst}
\end{figure}

Taking this effect into account, we investigated the implications of a constant accretion rate set to $\dot{m} = 0.1\,\dot{m}_{\rm edd}$ on the resulting long-term changes in QPE peak luminosities that are shown in Fig.~\ref{mdotconst}. Again, we used Eq.~\eqref{Lqpe} with the bolometric correction from Fig.~\ref{fig_QPE_SED}, which implies that $\kappa_{\rm bol}^{-1} =  L_{\rm x}/L_{\rm bol} \approx 0.9$. In contrast to the previous investigation, we considered the decreasing stellar mass due to ablation and, hence, the decreasing stellar radius as well. For simplicity, we assumed a stationary evolution along the MS and, hence, we obtained the corresponding radius-mass relation.

We find that the stellar ablation leads to a rather mild decrease in the QPE peak luminosity, as the shrinking stellar radius reduces the cross-section and, hence, the energy injection. For collisions with more abundant lower-mass stars, this decline remains small (at the level of a few percent) while for less abundant initially more massive stars, the relative X-ray peak luminosity can decrease significantly (e.g., $\approx 50\,\%$) for an $8\,\rm M_\odot$ star. The decline is the most profound at earlier times. However, deviations from the power-law radius-mass scaling are expected in realistic setups due to nonequilibrium evolution of the stellar structure \citep{Lu.et.al} since the thermal (Kelvin-Helmholtz) timescale of the star is much longer than the recurrence time between the collisions. Caused by heating of the upper stellar layers due to shocks, the stellar body is thus expected to initially expand rather than shrink. The detailed hydrodynamical treatment of the time-dependent stellar structure is beyond the scope of the current paper.

\section{Discussion}
\label{sect_discussion}

\subsection{Disk colliding with stellar debris}
As theories using the EMRI system as the main driver behind QPEs are  increasingly in development, a novel notion regarding the dominant source of the emission has emerged. At present, a scenario involving a crash of the star itself with the disk is a largely popular mechanism, but $M_1$ would end up losing a lot of material via ablation during this process. Analytical estimates made by \citet[][see their Fig. 11]{yao2024} show that the material released from the star forms a ``triaxial" ellipsoid. This elongated cloud of stellar debris is stretched along the orbit of the EMRI, piercing the accretion disk and subsequently creating shocks. The first analytical and numerical approximations of the eruption timings and emission were recently carried out in \citet{Linial_2025}, concluding that these estimates are aligned with current observational trends, such as $\Delta t \propto P_\text{QPE}$ and $E_\text{ej} \propto P_\text{QPE}$ (see also their Section 6.4). In other words, equations describing the emission and timings of stellar debris during the transition are aligned with a trend that supports the correlation of the transition duration and energy released with the QPE periods.

If the predicted mass of the debris stream ends up overcoming the mass of the disk it intercepts, strong shocks will occur, interrupting the azimuthal flow of the disk. However, the condition \cite[see Section 3.4 of][]{Linial_2025} for this kind of regime would imply a very small orbital period of the EMRI, which only three QPE sources would be able to satisfy for fiducial values ($M_1$ mass, stripping efficiency, etc.). Otherwise, the stellar debris would get deflected or mixed into the flow. This condition is an approximation, meaning that it would need to be tested via a hydrodynamical simulation.

\subsection{Context of partial TDEs}

To reproduce the timing properties of QPEs, the star has to orbit so close to the SMBH that it balances on the verge of its own destruction (i.e., close to the tidal radius). Scaling this effect to the typical QPE system parameters, it is convenient to establish the formula for the penetration parameter using Eq.~\eqref{beta} and 3rd Kepler's law via\begin{align}
    \beta = R_{\star}\Bigl(\frac{M_{\bullet}}{M_{\star}}\Bigr)^{1/3} a^{-1}(1-e)^{-1} \leq 0.4,\\
    P_\text{orb,d}^{-2/3} (1 - e)^{-1} M_\text{1}^{7/15} \leq 1 ,\\ \text{(disruption condition)},
    \label{impact_par}
    \end{align}
where the inequality is given to ensure that there is no mass transfer in the EMRI; here, the borderline value $\beta$ for a partial TDE is taken from \citet{fine_line}. As we  only have crude estimates of the recurrence time values, all QPE EMRIs could be overflowing their Roche lobes. The most probable candidates (with the shortest eruption durations) are eRO-QPE2 ($\beta = 0.8$), RX J301.9$+$2747, and GSN 069 for the fiducial parameters, mainly on quasi--circular orbits.

Now we have two constraints on the EMRI systems: the first one follows from Eq.~\eqref{condition} and the second from Eq.~\eqref{impact_par}. We set these constraints into Table~\ref{tableconstraints}.

\begin{table}[htp]
    \centering
    \caption{\centering Narrowing down the parameter space of QPEs.}
\resizebox{\columnwidth}{!}{
\begin{tabular}{c|c|c|c}
    \hline
    \hline
    Source & Collision cond. & Disruption cond. & Implications \\ \hline
    GSN 069 & 1.9-2.2 & 1.2 & $\downarrow \beta$ \\
    eRO-QPE1 & 2.9-3.5 & 0.7 & $\downarrow \beta$ + $\uparrow M_\text{dis}$  \\
    eRO-QPE2 & 0.6-0.8 & 2.9 &  $\downarrow M_1$ \\    
    RX J1301 & 1.3-1.5 & 1.9 &  $\downarrow \beta$, $ \downarrow M_1$ \\
    Swift J0230+28 & 32-36 & 0.1 & long after $t_\text{vis}$ \\
    eRO-QPE3,4 & 3.0-3.7 & 0.7-1 &  $\downarrow \beta$ + $\uparrow M_\text{dis}$ \\
    eRO-QPE5 & 11-13 & 0.3 & long after $t_\text{vis}$ \\ \hline
\end{tabular}
}
\tablefoot{
Both conditions should yield a number $\leq 1$. In the rightmost column are the modifications that could correct the discrepancies - modifications point to smaller or larger values. The conditions follow from~\eqref{impact_par} and~\eqref{condition}. The ranges of values in column 1 come from the SMBH uncertainties.
}
\label{tableconstraints}
\end{table}

The two most often discussed scenarios where the star is known to be repeatedly disrupted are: 1) a MS star being disrupted and 2) white dwarf (WD) that is orbiting the SMBH on a highly elliptical orbit. Both of these scenarios require finetuning and largely improbable conditions \citep[see the discussion in][]{arcodia2024}. In the case of the WD, for example,  we would need a very specific pericenter distance.

These arguments are also valid for other long-term observed QPEs. Most importantly, the star becoming robbed of its shell gets a kick 
that will throw it out of its orbit \citep{Chen_2021}. The orbital period of the survived core will become $\sim 400-40\,000\,$yr. 

\subsection{A rebrightening event of the long-lived GSN~069}
The X-ray luminosities of GSN 069 evolve extremely slowly \cite[][see their Section 5]{guolo2025}. Over nine years of the observations, the blackbody temperature declined only by about 25$\,\%$ compared to other confirmed post-TDE systems, such as ``Ansky" and others \cite[e.g., see][]{ajay2024}. This is probably linked to the viscous timescale, which is the main parameter controlling the evolution of the disk. In the cited work, the X-ray luminosity light curves were fitted using the temporal surface density evolution equation. The $t_\text{vis}$ parameter from this fit is an order of magnitude or two higher than the $t_\text{vis}$ fits found for other TDEs \citep{guolo2025}.

Clearly, this does not stand in contradiction to  the conclusion given in Section~\ref{sec3}, stating that we should observe QPEs a few years after the circularization of the debris. As $t_\text{vis}$ is the free parameter, we chose it to be approximately equal to $T_\text{cir}$ so the factor $\mathcal{R}$ would be in better agreement with the amplitude decline in other QPEs. From the modeling of GSN 069, we have $t_\text{vis} \approx 2300\,$d, which indicates that we are seeing an amplitude evolution taking place on much longer timescales. However, the absence of QPEs in 2014 and their appearance in 2018 remains a puzzle that cannot be explained by simplified theoretical models such as ours.

\subsection{Cloud expansion}
The luminosity of a spherical expanding optically thick cloud can be estimated from the elementary equation, $L = 4 \pi R_{\rm cl}(t)^2 F_\text{BB}$. The two quantities in this equation that are both dependent on properties of the accretion disk are the cloud radius $R_{\rm cl}(t)$ and its temperature, $T$. Using the model described in Appendix~\ref{apendixb}, it would be more consistent to work with some theoretical prediction of $R_{\rm cl}(t),$ instead of using the observationally inferred value. If we use Equation (16) from \cite{emritde}, we have
\begin{equation}
    \label{diffusion radius}
    R_\text{cl} = v_\text{K} t_\text{QPE} \propto \dot{m}^{-1/2},
\end{equation}
where the long-term luminosity scales with $R_\text{cl}(t)$ as $L \sim \dot{m}^{-1}$. There is an additional nonanalytical scaling from the temperature that makes this decline in time more shallow. Unfortunately, in this case, the luminosity scales inversely with the accretion rate, which is in disagreement with Eq.~\eqref{Lqpe}. However, this discrepancy can be reconciled with the observations since the cloud is not likely to expand linearly (i.e., $R_{\rm cl} \propto t$) into a vacuum; however, we do expect to see a hot diluted ambient medium \citep{1981ApJ...249..422K} that slows down the expansion due to the ambient pressure, especially at later times. In the case of a slower cloud expansion, the theoretical long-term luminosity evolution can be reconciled better with the observed one.

\section{Conclusions}
\label{sec4}

In this work, we revisit the interpretation of a subclass of RNTs, defined as QPEs that share some similarities with other types of RNTs, such as repeating partial TDEs, quasi-periodic ultrafast outflow sources, and so on \citep[see e.g.,][]{2024SciA...10J8898P}. In particular, this includes QPEs exhibiting large-amplitude soft X-ray outbursts with a short recurrence timescales. After summarizing the basic properties of this class of RNTs and connecting it to post-TDE systems, we invoked a specific model of QPEs that includes the EMRI model complemented by a preceding TDE. 

We discuss an important aspect concerning the size of the outer disk radius, finding that the disrupted star $M_\text{dis}$ was probably destroyed farther from the SMBH, with a smaller $\beta$ ($\beta \sim 1$). This could also imply that the disk was formed by a partial TDE since $\beta$ is smaller in this case. Using the statistical properties and analytic relations of QPEs, we derived the asymmetric light curves expected to be caused by star-disk collisions in a post-TDE system. We also discuss how the relative position of the disk and the EMRI orbit impact the emerging QPE luminosities.

Following this setup, we consider the implications of the dependence $L_\text{char} \propto \dot{m}^{-p}$. This is a simpler description of the emission since many perturber--induced QPE models have not considered analytical approximations; rather a combination of analytical and numerical approaches have been applied instead, using, for instance, a temperature profile of the disk as the main parameter for calculating the typical luminosity profiles. Nevertheless, we see that to have a sufficient drop in the typical eruption luminosities while accounting for the adequate amount of time for the circularization of the debris, the amplitude drop in $\sim$ 2--4 years can be reproduced only if the starting observation window had occurred at a few $\times (10\text{--}100)\,\rm t_\text{fb}$. The ratio of the X-ray luminosity to the bolometric luminosity is found to evolve rather slowly with the accretion rate. The temporal evolution of the $\mathcal{R}$ ratio (ratio of luminosities $L(t)$ with the fixed time separation) is thus qualitatively the same for the X-ray luminosity. Secondly, we find that both the time evolution of the accretion rate $\dot{m}(t)$ and the surface density $\Sigma(t)$ impact the QPE luminosity since the disk size spreads over many orders of magnitude in the first decades after TDE. If we accounted only for one of these effects, the amplitude drop would not be sufficient and a different luminosity prescription would be needed.

Finally, following the most recent simulations, we highlight the importance of stellar ablation, which has a noticeable influence on the energetics of star-disk collisions, especially in the case of more massive MS stars. These material losses could also be significant in the case of a steadily accreting disk.

\begin{acknowledgements}
The authors thank the anonymous referee for constructive comments that helped to improve the manuscript.
MM, MZ, HB, and VK acknowledge the financial support of GA\v{C}R Junior Star grant no. GM24-10599M; PK was supported by the grant GAČR GF23-04053L, and TJ by Physics for Future – Grant Agreement No. 101081515.
\end{acknowledgements}


\appendix
\section{Statistics}
\label{apendixa}

In this part of the paper, we show supplementary visualizations of the main QPE properties in four plots. Horizontal lines mark the median values of the whole population. As we are navigating through a few orders of magnitude, a median was chosen instead of the classical arithmetic mean. Lastly, a ratio of $\sigma_+/\sigma_-$ is shown for eRO-QPE1.

\begin{figure}[htp]
    \centering
    \includegraphics[width=0.8\linewidth]{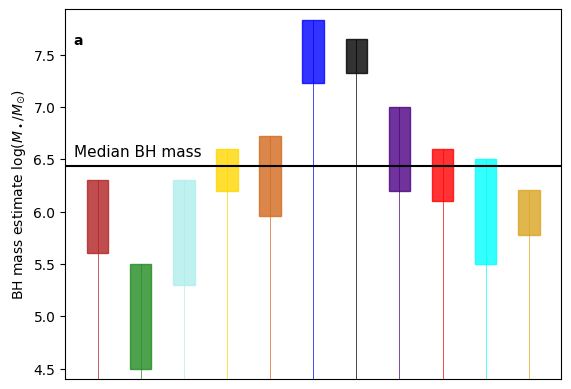}
    \includegraphics[width=0.8\linewidth]{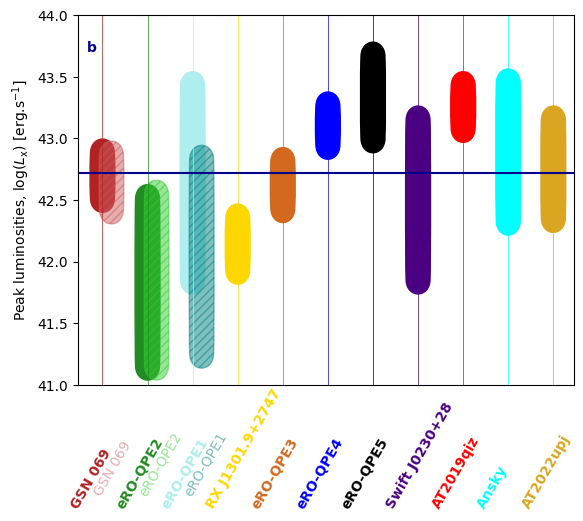}
    
    \caption{QPE properties 1. Light-dark color rectangles represent the time evolution of the features. 
     Light - first epoch of observation. Dark shows last epoch of observation. Every shape has the same width. Top panel: SMBH masses, mostly based on the $M-\sigma$ relation. Bottom panel: Inferred luminosities, based on the best spectral fit results of the accretion disk model.} 
     \label{properties1}
\end{figure}

\begin{figure}[htp]
    \centering
    \includegraphics[width=0.8\linewidth]{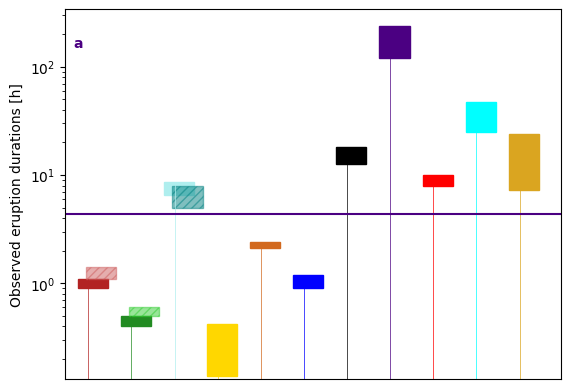}
    \includegraphics[width=0.8\linewidth]{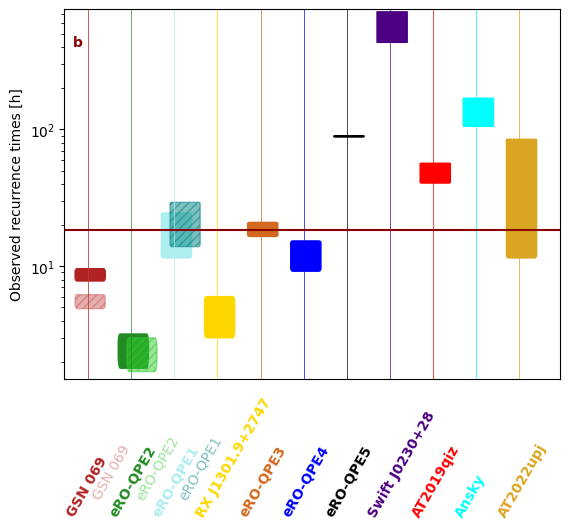}
    \caption{QPE properties 2. Top panel: Durations of the QPE eruptions. Uncertainties are not significant. The dispersion of individual eruption durations is, though. Bottom panel: Same as the top, but for the time intervals between individual eruptions.}
    \label{properties2}
\end{figure}

\begin{figure}
    \centering
    \includegraphics[width=0.7\linewidth]{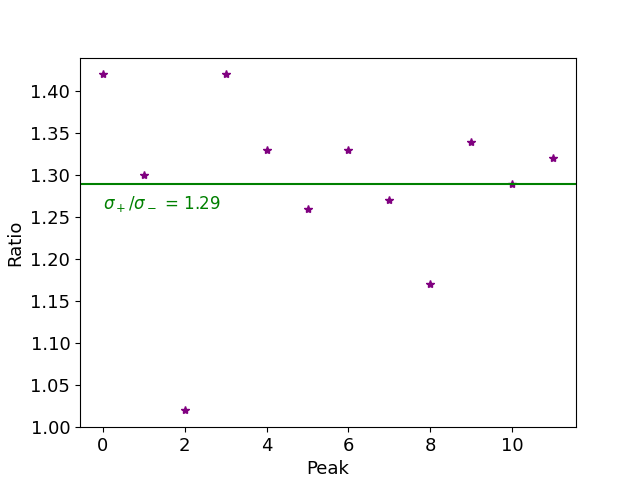}
    \caption{Decline and rise times for multiple eruptions in eRO-QPE1 from the epoch of 19th August 2020. The green line represents the average. The values of $\sigma_+$ and $\sigma_-$ were obtained by fitting the eruption count rates by an analytical function referenced in Sect.~\ref{modeling of LCs}. The count rates were kindly shared by R. Arcodia.}
    \label{asymmetry}
\end{figure}

\section{$L_{\rm x}/L_{\rm bol}$ ratio}
\label{apendixb}

To quantify  the expected ratio $L_{\rm x}/L_{\rm bol}$,  we consider a simplified model of the star-disk interaction, where the star orbits the SMBH on a circular orbit at the distance $r$. For basic estimates, we adopt $P_{\rm QPE}=24$ hours, which for $M_{\bullet}=10^6\,M_{\odot}$ corresponds to the stellar orbit radius $r\simeq 315\,R_{\rm g}$. For a solar-mass MS star, this is outside its tidal radius of $r_{\rm t}\approx 47\,R_{\rm g}$. Hence, the star interacts with the disk twice per orbit and we assume a perpendicular interaction of the circular stellar orbit.

To mimic the temporal evolution of the disk accretion rate, we adopted a simple power-law evolution of its relative accretion rate $\dot{m}=\dot{m}_0(t/t_0)^{-p}$, where at the initial time $t_0 \sim 10\,\rm t_\text{fb}$, $\dot{m} \sim 1$. Furthermore, we set $p=1$, which simplifies the conversion between $\dot{m}$ and time, since for $t=10t_0$, $\dot{m}\sim 0.1$ etc.

\begin{figure}
    \includegraphics[width=\columnwidth]{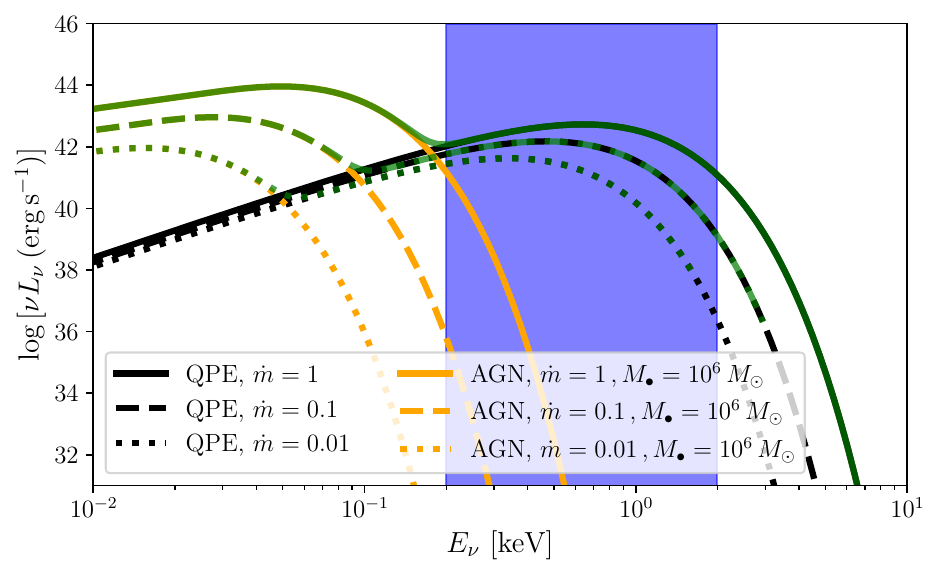}
    \includegraphics[width=\columnwidth]{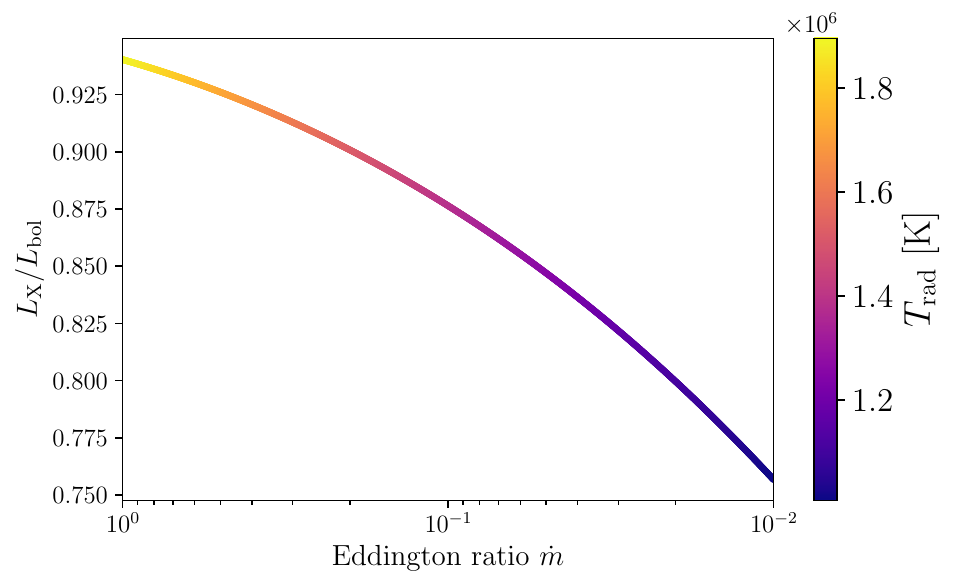}
    \caption{Evolution of the QPE spectral energy distribution with the relative accretion rate. Top panel: Blackbody spectra of ejected plasmoids (black lines) for different relative accretion rates ($\dot{m}=1-0.01$), which influence the post-shock radiation temperature. The blue shaded rectangle depicts the spectral range between 0.2 and 2 keV. For comparison, we also depict corresponding AGN spectra (standard accretion disks, orange colors) and their sum with QPE spectra (green lines). Bottom panel: Dependence of the ratio between the X-ray (0.2-2 keV) luminosity and the bolometric luminosity, $L_{\rm x}/L_{\rm bol}$, as a function of the relative accretion rate. The curve is color-coded using the post-shock radiation temperature.}
    \label{fig_QPE_SED}
\end{figure}

For calculating disk properties, we use the standard gas-pressure dominated accretion-disk solution, for which $\rho_{\rm d}\propto\alpha^{-7/10}\dot{m}^{11/20} M_{\star}^{5/8}r^{-15/8}$ \citep{1973A&A....24..337S}, where $r$ is the radial distance. The shock due to the stellar collision with an optically thick disk can be considered to be radiation-mediated, and the corresponding post-shock temperature can be derived using the equivalence between the radiation volume density and the ram pressure with a certain dimensionless efficiency factor, $f_{\rm rad}$,
\begin{equation}
    \frac{4\sigma_{\rm SB}}{c}T_{\rm rad}^4=f_{\rm rad} \rho_{\rm d}v_{\rm rel}^2\,
    \label{eq_QPE_temp}
,\end{equation}
where $\sigma_{\rm SB}$ is the Stefan-Boltzmann constant and $v_{\rm rel}$ is the relative velocity between the star and the disk material.
Hence, we expect that $T_{\rm rad}=[cf_{\rm rad}\rho_{\rm d}v_{\rm rel}^2/(4\sigma_{\rm SB})]^{1/4}\propto \rho_{\rm d}^{1/4}v_{\rm rel}^{1/2}\propto \dot{m}^{11/80}r^{-23/32}$. Since the left-hand side of Eq.~\eqref{eq_QPE_temp} is approximately constant for QPEs of different periodicities (distances from the SMBH) and relative accretion rates, we expect that $f_{\rm rad}\propto r^{19/8}$.
For $f_{\rm rad}=0.005$ ($\sim 0.5\%$ of the disk gas ram pressure is converted into QPE radiation energy density) and $P_{\rm QPE}=24$ hours, the radiation temperature is comparable to the one inferred from the QPE soft X-ray spectra, $T_{\rm rad}\sim 1-2 \times 10^6\,{\rm K}$. Subsequently, to obtain the X-ray luminosity in the range (0.2,\,2) keV and the total bolometric luminosity, we assume that the emission is thermalized, i.e., as implied by observational data, with the characteristic photosphere radius, which we set to $R_{\rm cl}=0.5\times 10^{11}\,{\rm cm}$ that yields the soft X-ray luminosity in the range $L_{\rm x}\sim 10^{42}-10^{43}\,{\rm erg\,s^{-1}}$; see Fig.~\ref{fig_QPE_SED} (left panel) for the example QPE blackbody spectral energy distributions. For comparison, we also show standard accretion disk spectra in AGN (orange lines), which are systematically shifted to lower energies.

In Fig.~\ref{fig_QPE_SED} (right panel), we show the ratio $L_{\rm x}/L_{\rm bol}=\kappa_{\rm bol}^{-1}$ (inverse value of the bolometric correction) as a function of the Eddington ratio and color-coded with the radiation post-shock temperature. We see that for the range $\dot{m}=1-0.01$, the ratio $L_{\rm x}/L_{\rm bol}$ evolves from 0.94 to 0.76 ($\sim 19\%$ drop) as the radiation temperature decreases from $\sim 1.9 \times 10^6\,{\rm K}$ to $\sim 1.0\times 10^6\,{\rm K}$.

\end{document}